\documentclass[review]{elsarticle}
\usepackage{setspace}
\usepackage{graphicx}
\usepackage{amsmath}
\usepackage{float}
\usepackage{hyperref}
\usepackage{cleveref}

\newcommand{\1}{$\langle100\rangle$}
\newcommand{\2}{$\langle110\rangle$}
\newcommand{\3}{$\langle111\rangle$}

\title{Stability of \1 Dislocations formed in W Collision Cascades}

\author[addbarcv]{Utkarsh Bhardwaj}
\ead{butkarsh@barc.gov.in}

\author[addhelsinki,addaalto]{Andrea E. Sand}
\ead{andrea.sand@aalto.fi}

\author[addbarcv,addhbni]{Manoj Warrier}
\ead{manojwar@barc.gov.in}

\address[addbarcv]{Computational Analysis Division, BARC, Vizag, AP, India-530 012}
\address[addhelsinki]{Department of Physics, P.O. Box 43, FI-00014 University
of Helsinki, Finland}
\address[addaalto]{Department of Applied Physics, Aalto University, FI-00076 Aalto, Espoo, FINLAND}
\address[addhbni]{Homi Bhabha National Institute,
Anushaktinagar, Mumbai, Maharashtra, India - 400 094}

\begin{document}

\begin{abstract}
 Experiments and simulations both have verified the presence of \1 dislocations 
 in irradiated W. It is essential to know the properties and behavior of these 
 defects to study the evolution of microstructures at higher scales. We study 
 the thermal stability and transition mechanism of various \1 dislocations 
 formed in a molecular dynamics (MD) database of 230 collision cascades using 
 three different interatomic potentials. The activation energy to transition 
 to more stable \3 dislocations is found for various \1 dislocation defects
 that transition within the 100 nanosecond time scale that is readily
 accessible to MD. The stability of \1 dislocations increases with size, but
 the trend is not strict. The reasons for irregularities are the aspects of
 internal configuration such as (i) the arrangement of \1 directed crowdions
 within the defect, (ii) the presence and arrangement of non-\1 crowdions on
 the fringes of the defect. We show the typical pathways of transitions and
 discuss the sources of instability in the defect configurations. We also
 discuss the similarities and differences in stability found across different
 interatomic potentials. Understanding transition mechanisms and internal
 morphology gives insights into the stability of \1 dislocations, useful in
 higher scale models such as Kinetic Monte Carlo (KMC).

\end{abstract}

\maketitle

\section{Introduction} \label{sec:intro}

Molecular dynamics simulations of collision cascades have been used extensively
to determine the radiation damage at atomistic scales \cite{StollerRadDamage}.
The higher scale models of irradiation induced changes in microstructure such
as KMC require defect distribution and their properties like diffusion,
stability and interaction as inputs \cite{WirthMDKMC, BecquartMDKMC,
StollerMDKMC}. Traditionally, the initial defect distribution input from MD
simulation consists of the number of point defects and defect cluster size
distribution. These defects have different morphologies which may have
different properties for diffusion, stability and interaction. For instance, in
W, two kinds of loops are observed: 1/2 \3 loops which easily glide in one
dimension and \1 loops which glide very slow, and then there are rings that are
completely sessile \cite{PhysRevMaterials.5.093605, savi, potcmp}. For
simplicity, in various KMC studies, point defects are considered glissile while
all other clusters, or clusters above a certain size, are considered sessile
\cite{lakimoca}. The diffusion affects the extent of interaction between the
defects which decides long term microstructural changes. It has been shown that
such assumptions about the transport properties has an affect on the predicted
damage results of the KMC simulation (\cite{becquart:hal-03011208} and
references therein). It is also important to track morphological transitions
of defects because transition to a different morphology will change further
diffusion and interaction of the defect. 

There have been various studies investigating the morphological distribution of
defects at different PKA energies in W \cite{potcmp, savi, SETYAWAN2015329, Sand_2013}. The
results show that while majority of the SIA defects consist of \3 dumbbells and
dislocation loops, there are also \1 loops, mixed loops, C15 like rings and
their basis structures, and mixed rings and dislocation loops. The \3 dumbbells and
their clusters forming \3 dislocations are the most stable self-interstitial
atom (SIA) defect morphologies in W. These constitute the majority of defects
formed in W collision cascade simulations \cite{Sand_2013, SETYAWAN2015329,
potcmp}. The prevalence of \3 dislocations in W simulations is in agreement
with the experiments \cite{yi2012, Yi15, durrschnabel2021new}. In addition to
the \3 dislocations, formation of \1 dislocations on high energy irradiation of
W has also been reported in experiments \cite{durrschnabel2021new, Yi15,
yi2012} as well as in MD simulations of collision cascades \cite{Sand_2013,
SETYAWAN2015329, savi}. The \1 loops inhibit void nucleation and affect the
irradiation swelling resistance of the material \cite{little1980swelling}.
Therefore, it becomes important to account for the \1 dislocation loops and
their corresponding properties in higher scale modeling of radiation damage in
W. 

The \1 dislocations have been found in MD simulations carried out with
various widely used interatomic potentials \cite{SETYAWAN2015329, potcmp}.
Different potentials show differences in numbers and sizes of different defects
morphologies \cite{potcmp}. The PKA energy dependence on morphology of defects
has also been studied \cite{SETYAWAN2015329, savi, potcmp}. The contribution of
bigger mixed loop defects increases after around 100 keV while the
contributions of single loops specifically \1 loops decrease. However, the
maximum size and average size for \1 loop increases with energy for the range
10 keV to 200 keV. 

There have been MD studies to explore the stability of
specific defects. In \cite{GAO2000213}, stability of three defects of size 2, 8
and 13 produced by displacement cascades in Fe is studied. The three defects
studied have different morphologies and sizes. The defects are first isolated so
that their stability can be studied without interaction and influence
of nearby defects. The defects are annealed at different temperatures and their
transition times are noted to find the activation energy for transition
(transition energy). Four sample runs are given for size-2 defect at five
different temperatures while a single run is given for other two defects at four
different temperatures. The maximum duration of a single MD simulation is 1 ns.
In \cite{BONNY2020109727}, transformation of sessile multi-dislocation defects
to single 1/2 \3 dislocation is reported by annealing primary damage state of
ten 200 keV cascades. Each cascade is annealed at a single temperature of 1500
K for up to 5 ns. Three different interatomic potentials are used to study the
effects of interatomic potentials. The estimated transition energy for these
defects is reported to be approximately 1.5 eV. As noted in the study, the
transition energy is only indicative or qualitative because of the presence of
other defects that affect stress fields and may interact with the defect being studied.

The shape and internal configuration of \1 loops has also been studied. Eyre and Bullough in \cite{rectilinear100} show that the experimental findings of rectilinear shape
of \1 loops in bcc metals is due to lower elastic energy of
rectilinear shape when compared to circular. The \1 loops maintain
rectilinear shape as the size grows while bigger \3 loops are
circular. The study also shows that the rectilinear loops will tend to
be square i.e. the ratio of length to breadth (aspect ratio) of the
rectilinear shape will tend to be lower. Setyawan et el. in
\cite{SETYAWAN2015329} show that the shape of \1 loops observed in MD
simulations of collision cascades in W are parallelogram (or a
rhombus). It has been shown for \3 dislocations in Fe that the energy
density of the crowdions in the center of the defect is lower than
that on the interface \cite{Dudarev111clusters2003}. This has also
been observed and analyzed for both \3 and \1 dislocations in W
\cite{savi}. The study also shows that the \1 dislocation loops found
in MD simulations of collision cascades in W have a \1 bunch of
dumbbells surrounded by non-\1 dumbbells on the fringes.

This work examines the thermal stability and transition mechanism of \1
dislocations of varying sizes and configurations formed in W collision cascade
simulations using MD. We examine 34 \1 dislocations in a database of 230
collision cascades carried out using three widely used interatomic potentials.
We isolate the defects and then carry out annealing MD simulations. To
calculate transition energy of a defect, we carry out 16 runs for maximum of 10
ns at 12 different temperatures which amounts to 192 runs to estimate the
transition energy of a single defect. We carry out a systematic study with size
and discuss the relationship between size and transition energy of \1
dislocations.  The activation energy for transition to \3 dislocations of
various medium-size \1 dislocations, including clusters of 6 to 24 self
interstitial atoms (SIAs) is found to be in the range of 0.1 eV to 3.2 eV. We
show that the transition energy increases with an increase in defect size.
However, the trend is neither continuous nor regular. To understand the reason
behind irregularities, we analyze the internal configuration of defects using
the SaVi \cite{savi} algorithm. We show that the differences in transition
energy of same sized defects is due to the differences in their internal
configuration while defects with different sizes also have similar stability if
the arrangement of \1 dumbbells is the same. We further understand the
relationship of internal configuration, defect size and defect stability by
showing the typical transition pathways that highlight the sources of
instability. We also discuss the similarities and differences in stability of
\1 dislocations appearing in the different interatomic potentials. 

\section{Methods} \label{sec:method}

\subsection{MD simulation}\label{sec:mdsim}

The \1 dislocations explored consist of 34 defects formed in 230 collision
cascades. The procedure used for the collision cascade simulations and the
identification of defect morphologies is discussed in detail in \cite{savi,
potcmp}. The database consists of collision cascades using three interatomic
potentials: Finnis-Sinclair potential \cite{doi:10.1080/01418618408244210} as
modified by Juslin et al. (JW) \cite{JUSLIN201361}, the potential by Derlet et
al. \cite{PhysRevB.76.054107} with the repulsive part fitted by Bj\"orkas et
al. \cite{BJORKAS20093204} (DND-BN), and the potential by Marinica et al.
\cite{Marinica_2013}, stiffened for cascade simulations by Sand et al. (M-S)
\cite{SAND2016119}. Collision cascades from multiple potentials help in
analysing the effect of choice of potential on the stability of \1 dislocations
loops. We selected these three potentials owing to the large collision cascade
database that is available with us for these three potentials.

The stability analysis and transition energy calculation is carried out on
the simulation results of several annealing MD simulations. The simulations were performed for each
\1 dislocation loop in isolation at different temperatures. We first extract
the desired defect along with five extra unit cells around it from the
simulation box of the collision cascade. We then add more unit cells of W on
each side of the extracted volume to form a simulation box of at least five
times the extracted number of unit cells in each direction to take care of
finite size effects. The Large-scale Atomic Molecular Massively Parallel
Simulator (LAMMPS) \cite{lammps} code is used to carry out MD simulations to
relax the system using an NPT ensemble at 300 K with periodic boundary
conditions (PBC) and zero pressure. This relaxation step is essential given
that we have placed the defects and their nearby unit-cells into a new crystal.
We analyse the dislocation after the relaxation for any change in structure or
inconsistency. 

For each relaxed system we carry out a temperature ramping simulation from 300K to 2000K. The temperature at which the defect transitions to \3 dislocation ($T_r$) in a ramping simulation is noted. We calculate 12 temperatures ($T_1$ to $T_{12}$) around $T_r$ with a difference of 25K i.e. $(T_2-T_1) = (T_3-T_2) = $ ... $=(T_{12} - T_{11})$ = 25K and $T_7 = T_r$. After that, sixteen different NPT runs are carried out at each of these
temperatures with different random number seeds for initializing the
temperature. Periodic boundaries are set at zero pressure. These simulations
are stopped either at 50 ns or if the defect transitions to \3 dislocation. For
cases where most of the sample runs transition while a few do not transition at
a certain temperature, the non-transitioning simulations are extended for 100
ns. The temperatures where transitions occur are included for the transition energy
calculation. We do not calculate the transition energy of a defect if it does
not transition for the complete 100 ns run even at a high temperature of 2000
K. We use an output frequency of 2 ps for checking transitioning and
configuration changes.

\begin{figure}[ht]
  \centerline{\includegraphics[width=.95\linewidth]{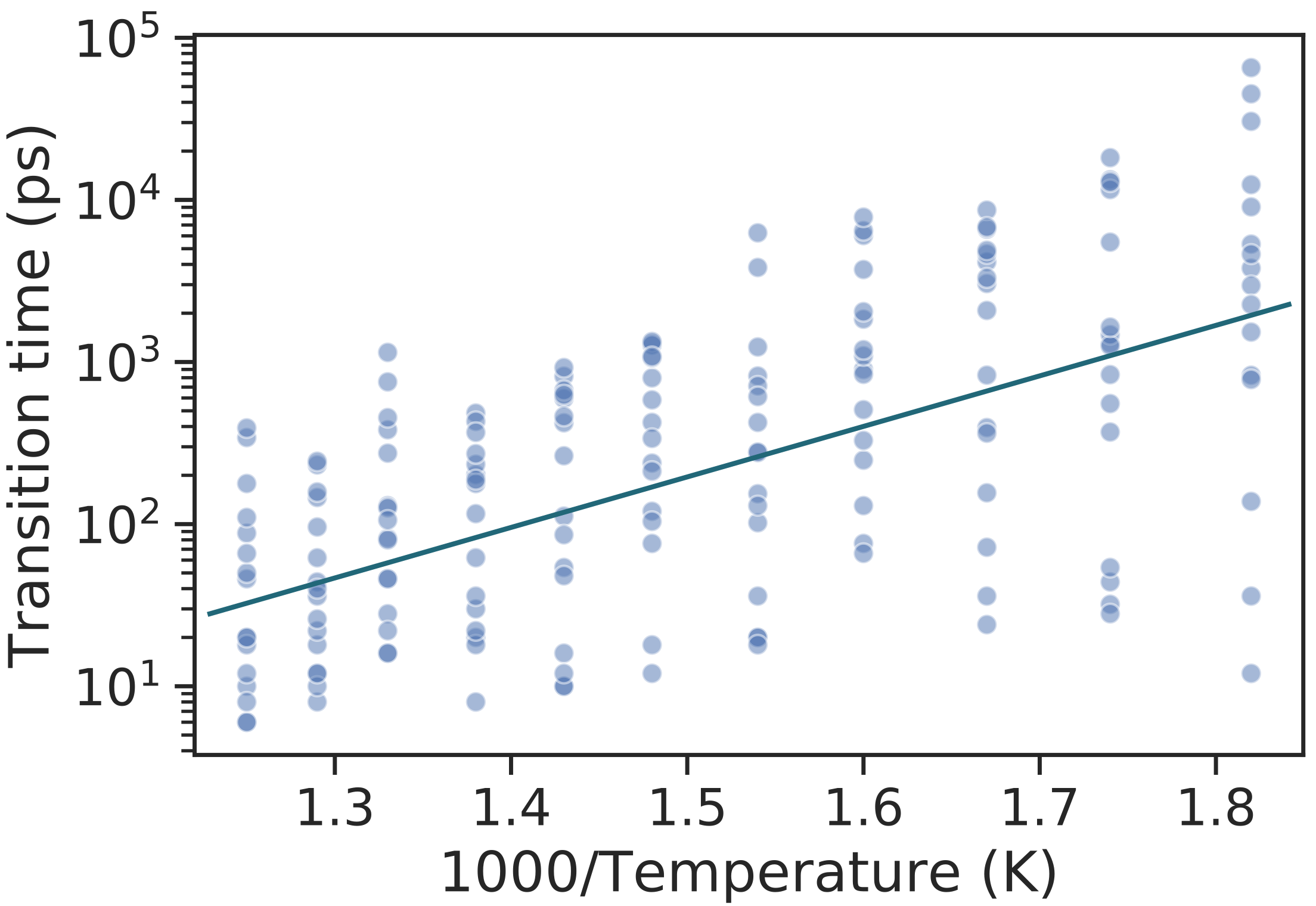}}
  \caption{\label{fig:fig1}
The transition time in various sample simulation runs of a size 11 \1 dislocation loop as a function of annealing temperature. The slope of the linear regression fit is proportional to the transition energy $E_t$ as given in \Cref{eq:arrh}.
} 
\end{figure}

For transition energy calculation, the temperature (T) dependence of time 
for transition ($\tau$) is taken to be described by an Arrhenius expression:

\begin{equation}
\label{eq:arrh}
\tau = \tau_o exp(E_t/k_b T)
\end{equation}

where $\tau_0$ is a pre-exponential factor, $E_t$ is the activation energy for transition to \3 loop and $k_b$ is the Boltzmann constant. A linear regression fit for the equation is found for different runs of each defect. \Cref{fig:fig1} shows a sample fit for a defect of size 11. The plot shows the time of transition of the \1 loop to \3 loop for different runs at eleven different temperatures. Out of the total twelve temperatures considered initially, for the lowest temperature a few of the runs did not transition within the 100 ns time limit. For this reason only eleven temperatures are shown in the plot. The temperature range is from 550K to 800 K with an interval of 25K. At each temperature there are sixteen different runs having different transition times. All these data-points from the eleven temperatures are included for the transition energy calculation. The transition time is governed by a Poisson distribution which results in a spread of transition times at each temperature. Moreover, different transition pathways and changes in configurations (as discussed in \Cref{sec:transmech} and \Cref{sec:internalConfig}) further add to the variation in the transition times of different runs at a specific temperature.

%\subsection{Transition energy calculation}

\subsection{Analysis of MD simulation} \label{sec:analMD}

We track morphological changes using the SaVi algorithm \cite{savi} and use
CSaransh \cite{bhardwajcsaransh} to visualize the defects. In addition to
identifying when the morphological transition occurs, SaVi also outputs various
parameters such as the number of dumbbells/crowdions in specific orientations
at each step and the number of neighbors of each dumbbell/crowdion. These
parameters define the internal morphology and help understand the relationship
between stability and internal configuration.

SaVi algorithm uses computational geometry and graph data-structure to find the
various morphological components of a defect and identify the overall
morphology based on the components. The algorithm finds lines across all the
dumbbells and crowdions. The direction of the lines decide the orientation of a
dumbbell or a crowdion. The lines that share a specific angle and distance form
a morphological component. For instance, a dislocation is formed by lines that
are approximately parallel to each other, while a hexagonal ring is formed by
lines that have 60 degree angle. The direction of the Burgers vector for a 
dislocation loop is found by the orientation of the constituent lines while the
magnitude is decided by the number of net defects in each line. The algorithm
involves several details that make it robust to noise such as finding cycles
for asserting rings. The algorithm can be used to find if a dumbbell is at the
center of the defect or on the surface by counting its number of nearest
neighbouring parallel lines.

\subsection{Naming Internal Configuration}

The arrangement of dumbbells and crowdions in a dislocation decides
the energy density and stability\cite{Dudarev111clusters2003,
rectilinear100}. This forms the basis of examining relationship
between stability and internal configuration. For the discussions of
internal configurations, it is important to express 
it in a way that conveys the shape of the defect especially
the factors that are observed to affect the stability.

\begin{figure}[ht]
  \centerline{\includegraphics[width=.95\linewidth]{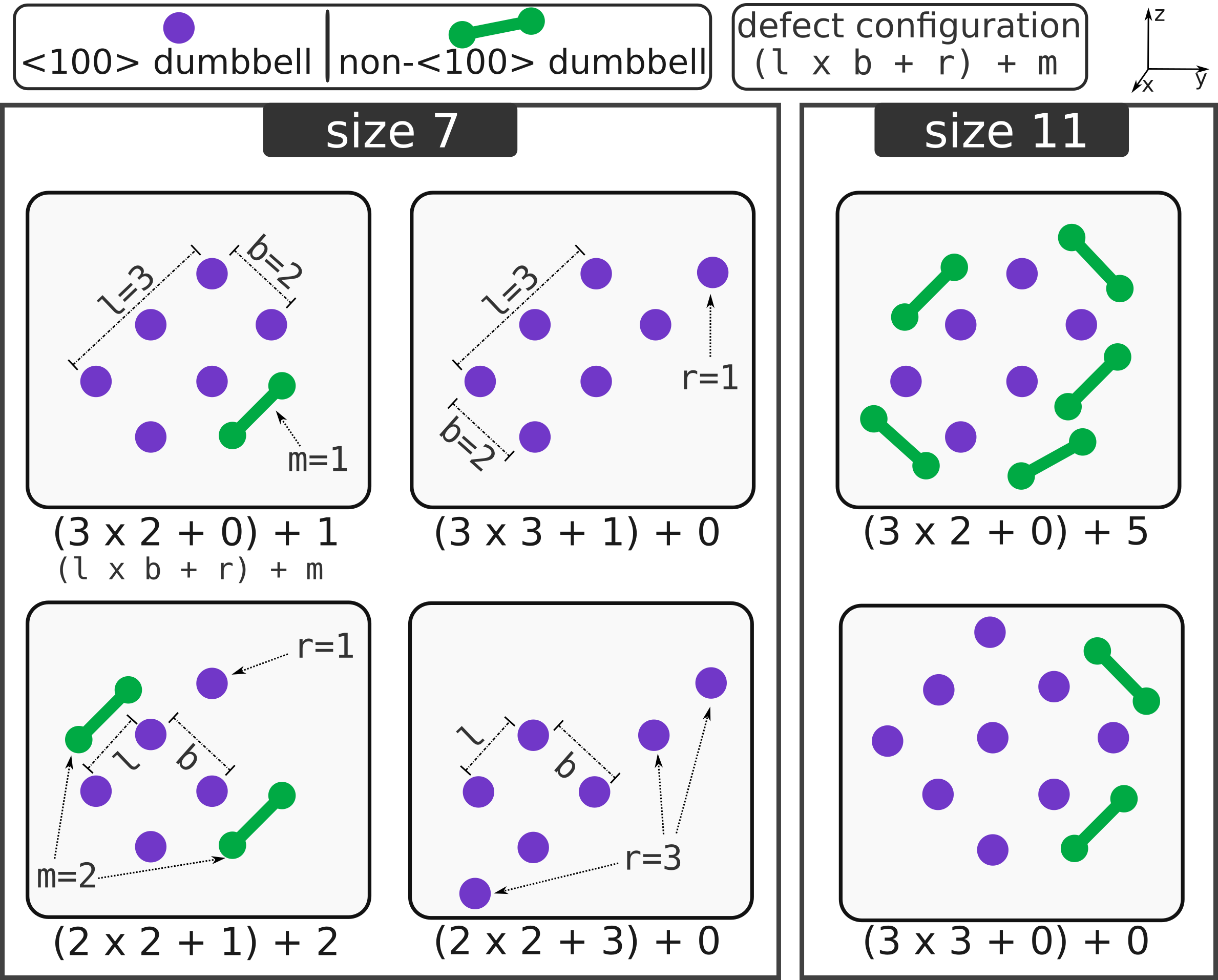}}
  \caption{\label{fig:fig2}
  The schematic shows naming parameters for four configurations of size-7 and
two configurations of size-11 \1 dislocations. The naming helps in expressing
the arrangement of dumbbells in a \1 dislocation. The parameters like $l$, 
$b$, their ratio also give an idea of the number of dumbbells in the center of
the defect and on the edges which has an affect on the stability of the
defect. The legend at the top shows colors and symbols for dumbbell orientation,
parameters for the expressions written below each defect and axis orientation
for the plot.
} 
\end{figure}

The \1 dumbbells in a \1 loop arrange in rectilinear form
\cite{rectilinear100}. More specifically, in a parallelogram or rhombus shape
\cite{SETYAWAN2015329} along with a few residual dumbbells outside the full
parallelogram \cite{savi}. We utilize the length, breadth and residual number
of \1 dumbbells to name the configuration of \1 dumbbells. The naming scheme
also signifies the degree to which dumbbells are packed at the center in a
configuration and can help in understanding stability differences for same
sized defects. The scheme used to name a configuration of size $s$ is as
follows:

\begin{equation}
\label{eq:eq1}
s = n + m
\end{equation}

\begin{equation}
\label{eq:eq2}
n = (l \times b + r)
\end{equation}

where, $n$ is the number of \1 dumbbells while $m$ is the number of non-\1
dumbbells on the fringes. The configuration of \1 dumbbells is further divided
into $l$, $b$ and $r$ where $l$ and $b$ are the length and breadth of the
number of completely filled rows and columns (sides of parallelogram) formed by
the \1 dumbbells. $r$ is the remaining or residual number of \1 dumbbells. The
values of $l$ and $b$ are to be chosen such that the residual $r$ has a minimum
value. \Cref{fig:fig2} shows a schematic representation of a few defect
configurations and their corresponding parameters for naming.

We say that two defect configurations are different if any of the values of
$l$, $b$, $r$ or $m$ are different. Another feature that we will note in a
defect configuration is the ratio of $l$ and $b$, which we refer to the aspect
ratio of the \1 component. This ratio together with number of residual dumbbells
gives an idea about the number of dumbbells that will be on the edges. As shown
in \cite{rectilinear100}, the \1 loops tend to have lower aspect ratio.

In place of current naming scheme, one can find the number of \1
neighbors for each \1 dumbbell and use this neighbour count for comparing the
stability of defects where sizes are same. A dumbbell in the center would have
four \1 nearest neighbors while the one on the interface will have less than
four. This scheme might be more appropriate for quantitative comparison or for
loops that are more often circular than rectilinear. One drawback of the
neighbor count based scheme is that it does not give a picture of the
arrangement for qualitative discussion. In \Cref{sec:internalConfig}, we will
discuss the relationship of the stability of a configuration with the different
factors of the notation that we have used.

\section{Results}\label{sec:res}

\begin{figure}[ht]
  \centerline{\includegraphics[width=.7\linewidth]{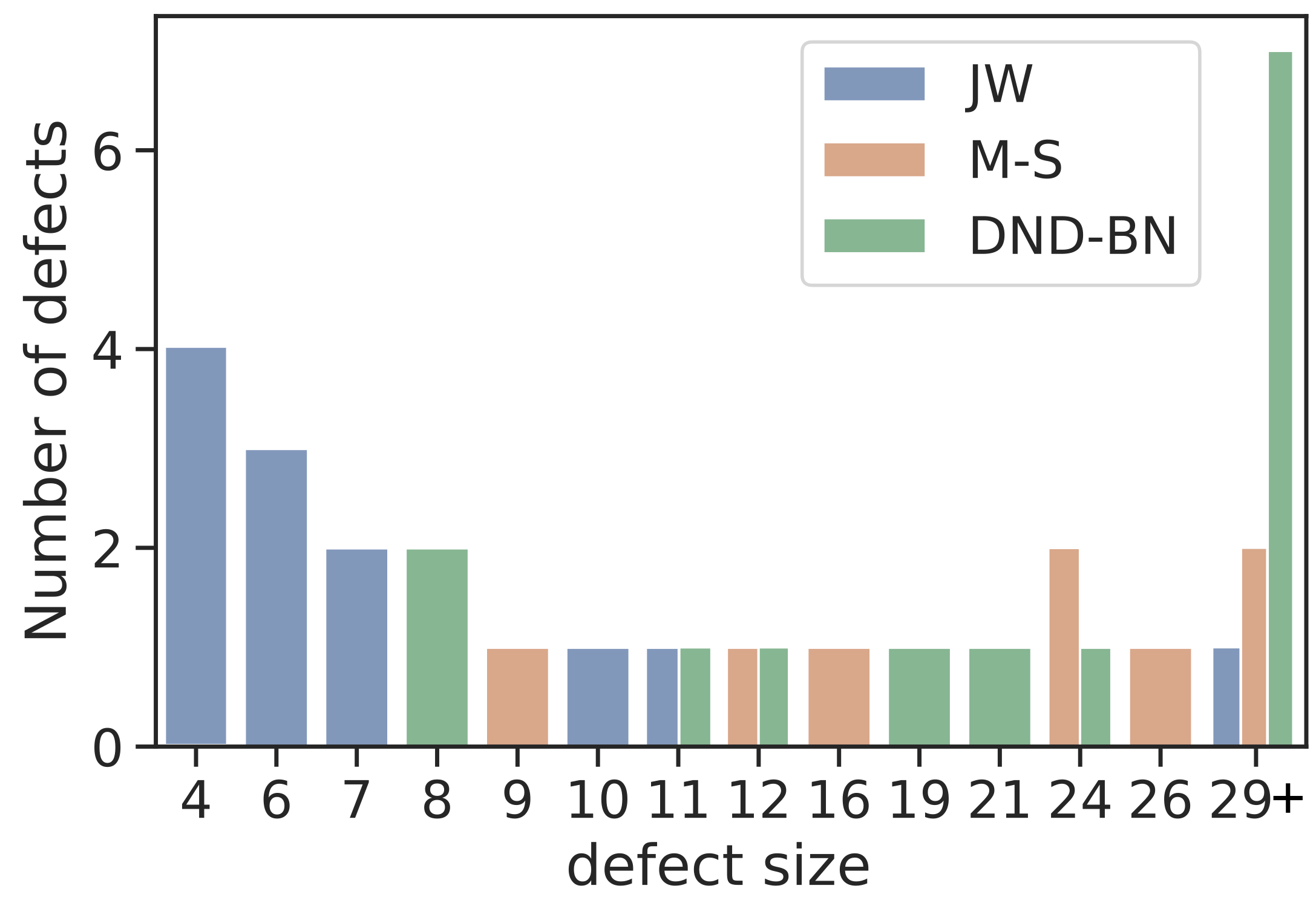}}
  \caption{\label{fig:fig3}
  The number of \1 defects for all the defect sizes present in the database for 
  each of the potentials. The values for each potential are shown separately 
  with different colors. Defect size implies the number of SIAs in the defect.}
\end{figure}

The database of \1 edge dislocations contains 34 defects found in a database of
230 collision cascades simulated with the three different potentials at PKA
energies ranging from 5 keV to 200 keV. The distribution of different
morphologies and defect size distribution for each morphology for a subset of
the database has been earlier shown in \cite{potcmp, savi}. The fraction of
defects forming \1 loops reduce after around 100 keV energy but the maximum
possible size of \1 loop increases with energy. There is no dependence of
internal configuration on PKA energy. All the three potentials predict damage
containing \1 dislocations, with almost all the defects having a size between 4
and 50 (the DND-BN potential has an outlier defect of size 147). The size of a
defect refers to the number of SIAs in the defect. \Cref{fig:fig3} shows the
frequency of different sizes of defects present in the database for each of the
potentials. Sizes 29 or bigger are grouped because there are no transitions
observed within MD timescales for defects of sizes 29 onward for all the three
potentials.

%fig-1

\begin{figure}[H]
  \centerline{\includegraphics[width=.99\linewidth]{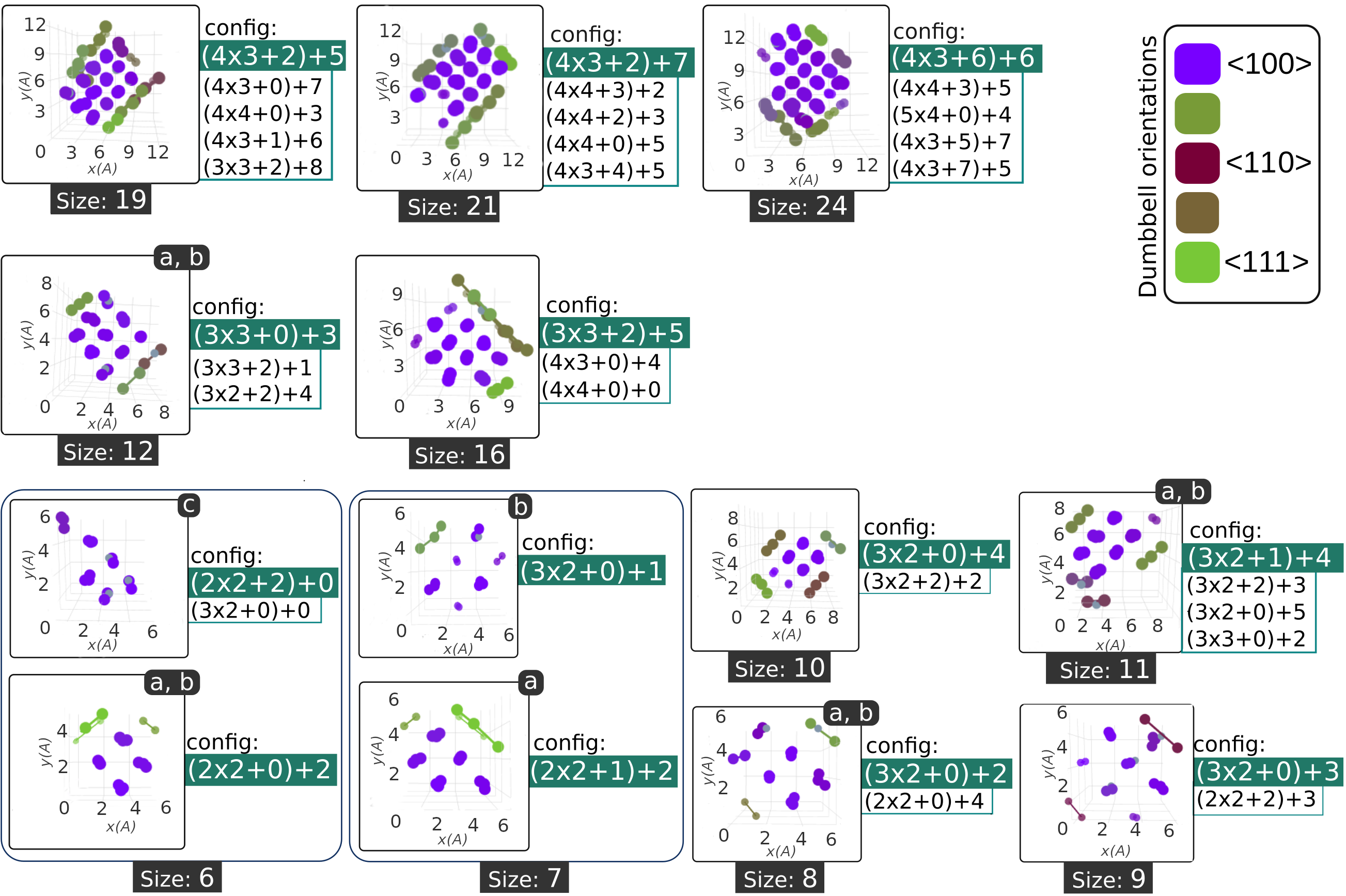}}
  \caption{\label{fig:fig4}
  Defects that show transition within the maximum MD simulation time of 100 ns
at a maximum temperature of 2000 K. A single configuration is shown for each
defect. Other prevalent configurations are listed for the defects that change
configuration. Different defects in same size are marked as a, b and c. A
single plot is shown for multiple defects if the defects have exactly the same
configuration such as for size 6 there are two defects with exactly same
configuration which are shown with a single plot marked with a and b. The axes
for each defect are shown to give an idea of the defect dimensions. 
}
\end{figure}

Out of the total 34 \1 dislocations in the database, the transition energies 
for 17 defects were calculated. From the remaining 17 defects, 13 large defects 
do not transition during MD simulations of 100 ns at 2000K temperature, whereas 
four defects of size four transition to \3 during the one nanosecond relaxation 
run at room temperature. \Cref{fig:fig4} shows the different configurations of
the 17 defects that transition to \3 dislocation. A single defect can
switch to different arrangements while being in \1 dislocation morphology
within a single simulation. The figure also lists other prevalent configurations for 
the defects having multiple stable configurations. 

\subsection{Configuration Dynamics and Transition Mechanism}\label{sec:transmech}

\begin{figure}[ht]
  \centerline{\includegraphics[width=.95\linewidth]{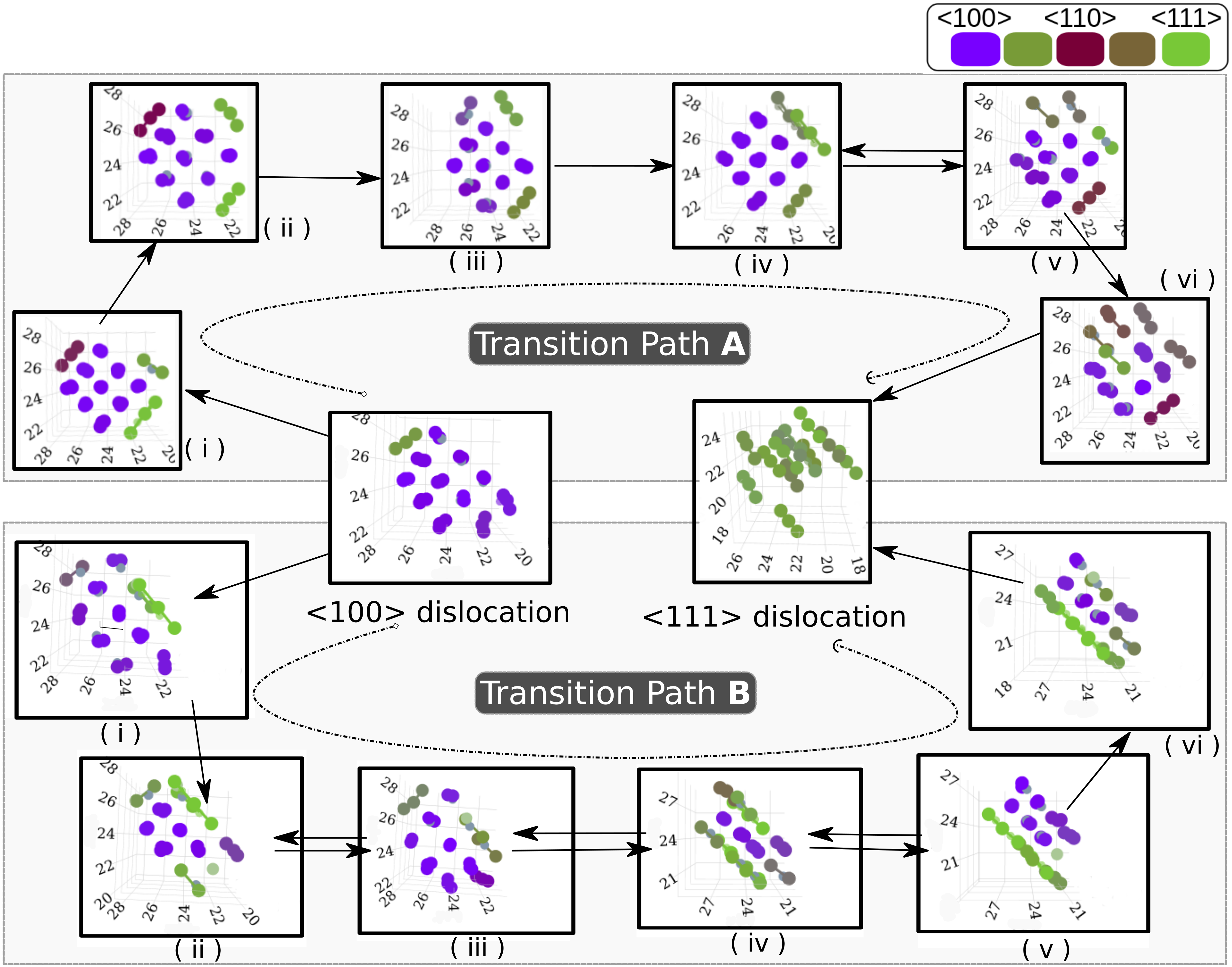}}
  \caption{\label{fig:fig5}
Two typical dynamics of dumbbell rotations and pathways for transition of
\1 dislocations to \3 dislocations. Path-A is common in configurations that
have no residual \1 dumbbells forming a \1 configuration with complete
parallelogram/rhombus. The arrows show the time evolution of configurations.
Different changes in configurations are shown as the size 12 defect transitions
to \3 dislocation. For both paths, the rotation of \1 dumbbells is initiated
at edges.
}
\end{figure}

We find the changes in configurations and transition mechanisms in the simulations by
tracing the movements and rotations of dumbbells that are in the center and on
the edges. Although in some cases the configuration changes can be too quick and
non-recurring, most of the times the transition pathways and configuration changes
fall clearly in one of the two categories shown in \Cref{fig:fig5}. These two paths are representative of the reoccurring patterns
that we observe for all the transitioning defects. The figure
traces the significant changes in defect configurations for two separate runs
of a size-12 defect as it transitions to \3 dislocation. The initial
configuration of the defect is $(3\times3+2)+1$ (shown left of the center in
\Cref{fig:fig5}). 

Path-A is the preferred path of configuration changes and transition when $r =
0$. In \Cref{fig:fig5} path-A starts when the defect transitions from initial
configuration to $3\times3+0$ configuration. The non-\1 dumbbells move around
the fringes from one side to the other while the \1 dumbbells remain stable in
the same configuration. This movement can result in a climb of the dislocation
loop, which is observed to occur rarely. However, glide in these dislocations
is much more common. As the non-\1 dumbbells move around, they may meet and
cluster, making the defect slightly unstable, inducing rotation in adjacent \1
dumbbells to non-\1. In our example, this is shown by \Cref{fig:fig5} (i) to
(v) of Path A. The rotation of the adjacent dumbbells in the \1 bunch may go
back and forth for some time. At some point, a good majority of the dumbbells
in the main bunch rotate, resulting in the transition to highly glissile \3
dislocation. In a bigger \1 dislocation, the number of \1 dumbbells is more,
and it becomes difficult to reach an instance when a majority of main \1
dumbbells are in the non-\1 direction. A bigger defect may either stay as a
multi-component mixed loop or may keep looping between stages similar to (iv)
and (v), thus taking longer to eventually transition.

Path-B is typical when $r > 0$ or multiple non-\1 dumbbells are distributed
unevenly around the \1 component. The exact arrangement of the
\1 dumbbells keeps changing very often as the \1 dumbbells on the
edges keep rotating from \1 to non-\1 as shown in \Cref{fig:fig5}
(ii) to (v). At a certain moment, the rotation may travel from the edge towards
the center of the dislocation. If a majority of dumbbells get rotated away from
\1, the dislocation rather than coming back to \1 sways entirely to \3
dislocation (\Cref{fig:fig5} (vi)). Again, if the value of $n$ is greater with
more \1 dumbbells packed inside, it is more likely for the rotating non-\1
dumbbells to sway back rather than the disturbance penetrating deep resulting in
the rotation of all \1 dumbbells.

\begin{figure}[ht]
  \centerline{\includegraphics[width=.95\linewidth]{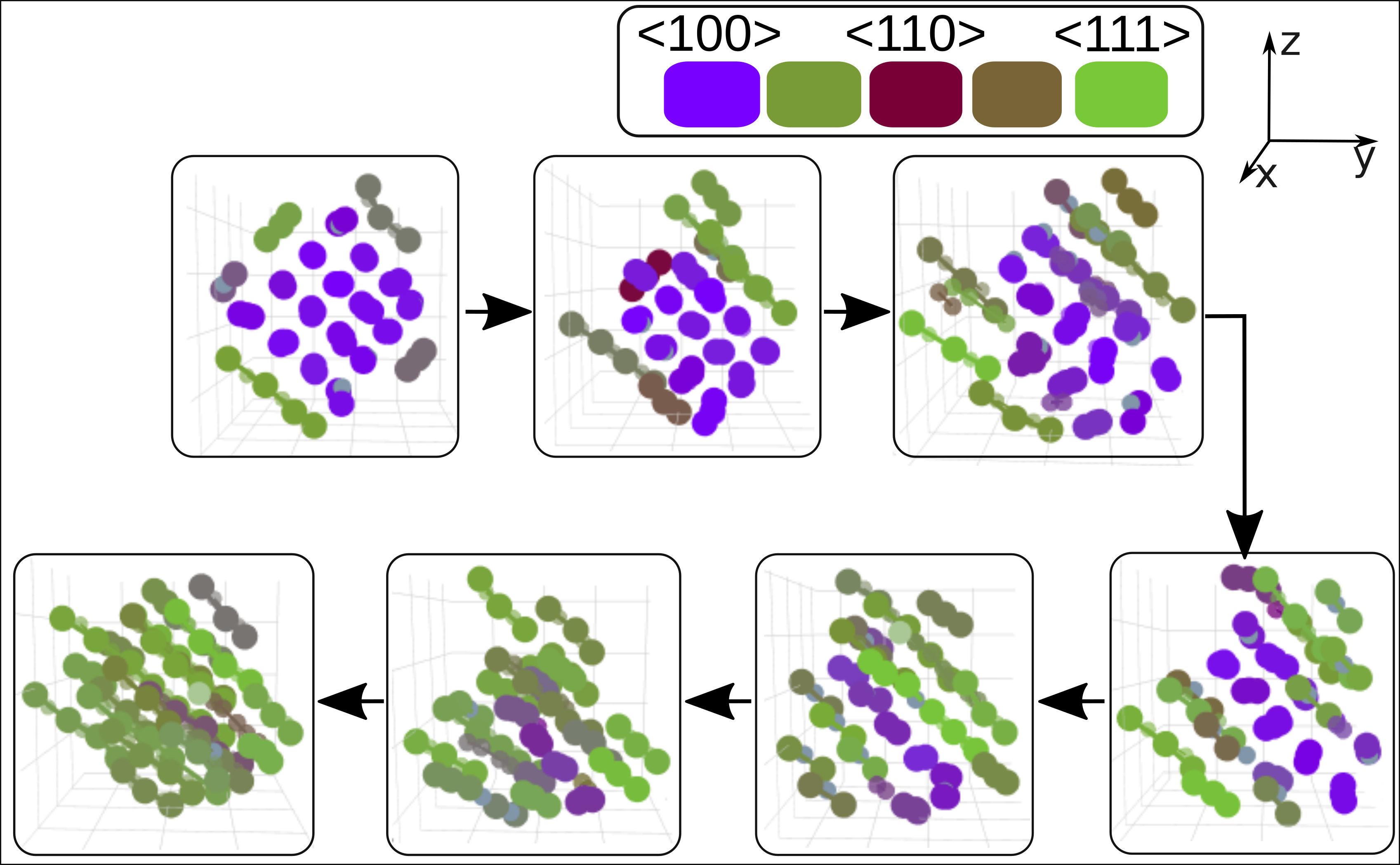}}
  \caption{\label{fig:fig6}
The transition pathway of a size 21 defect showing rotation of dumbbells from 
edges to center. As the initial configuration is a full parallelogram ($r=0$), path-A
is taken. The arrows show the time evolution of configurations. The configurations
in second row change relatively quickly.
}
\end{figure}

It must be noted that at any time the further dynamics of defect configurations may change from one path to another if the defect configuration changes in such a way. For instance, if an incomplete parallelogram ($r>0$) transitions to a smaller but complete parallelogram ($r=0$) then the pattern of further configuration changes will be based on path-A.

For both paths, the transition starts from the edges and moves towards the
central part. \Cref{fig:fig6} shows transition of a size twenty-one
defect. The rotation of dumbbells can be clearly seen to be initiated at the edges and then moving towards the center. For a bigger defect the half rotated configurations in first row are more stable. The defect may stay as a mixed loop for longer time. The configuration then may sway back to \1 again or may trickle down to \3 loop as in the second row of the \Cref{fig:fig6}. The final transition stages are quick and same in both the paths.The relative instability of dumbbells/crowdions at the edges compared to central ones can be understood by their higher energy density as shown in \cite{Dudarev111clusters2003, savi}. This difference in stability is also observed in our further examination of the relationship between internal morphology and stability in \Cref{sec:internalConfig}.

\subsection{Size dependence of Transition energy} \label{sec:sizedep}

\begin{figure}[H]
  \centerline{\includegraphics[width=.7\linewidth]{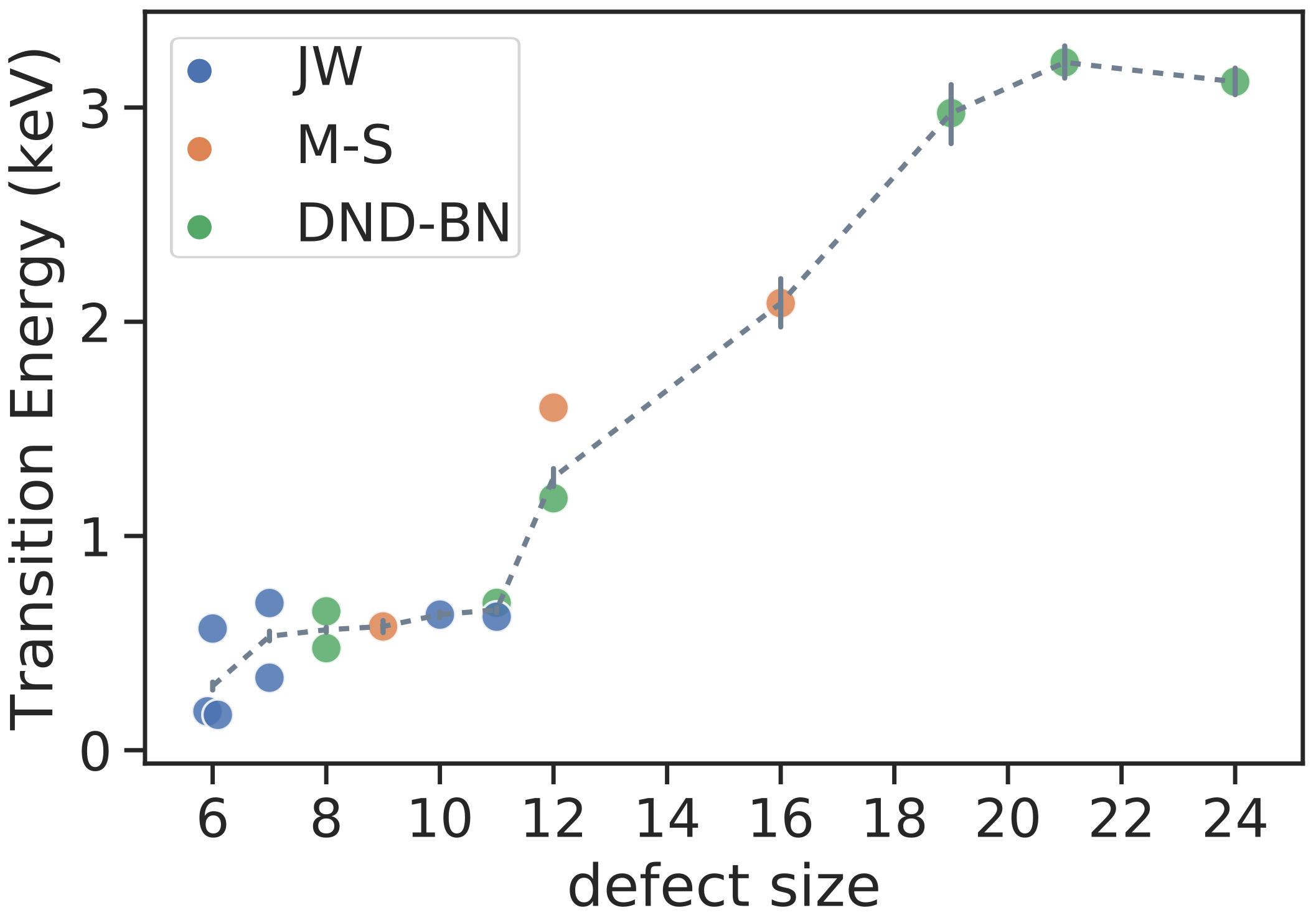}}
  \caption{\label{fig:fig7}
	Transition energy for various defects. Different colors are used for 
	different potentials. The transition energy increases with defect size. 
  The error bars represent the 95\% confidence interval for the transition
  energy fit.}
\end{figure}

\Cref{fig:fig7} shows the transition energy of various \1 dislocations for 
transitioning to \3 dislocations as a function of defect size. The
configuration of these defects are shown in \Cref{fig:fig4}. In case there are
multiple defects of the same size, the defects in \Cref{fig:fig4} are labeled
by alphabets (a), (b) and so on, starting from defect with lower transition
energy in \Cref{fig:fig7}. The overall trend in \Cref{fig:fig7} shows an
increase in defect stability with an increase in size. 

The majority of \1 dislocation loops that get formed in a collision
cascade are of the size ranges that will transition to \3 dislocation loops at
elevated temperatures. This is in agreement with the experimental findings that
the number of \1 loops reduce at elevated temperatures \cite{YI2016105}. 

The transition energy does not increase linearly with size but rather in steps.
It remains comparable for a few similar sizes and then increases sharply. For
example, in \Cref{fig:fig7} we see that sizes 7 to 11 have comparable
transition energy, and then for size 12, it increases sharply. Again, the
transition energy values are similar for sizes 19, 21, and 24, whereas size 29
is too stable to transition within the MD simulation time limits, even at a
temperature of 2000 K. We also observe that defects of the same size may have
different transition energy. For example, out of the three defects of size six,
two have similar transition energy, whereas the third has higher transition
energy (\Cref{fig:fig7}) though all three defects are created in collision
cascades using the JW interatomic potential. The internal morphology as discussed in next subsection reveals
the reason behind these observations.

\subsection{Internal morphology and stability\label{sec:internalConfig}}

\subsubsection{Transition energy of different configurations for the same size defects}

There are more than one defects for the sizes 6, 7, 8, 11 and 12. The transition energy of two defects with size 6 (labeled as (a) and (b) in \Cref{fig:fig4}) is same while another defect (labeled as (c)) has a higher value of transition energy. \Cref{fig:fig4} shows the difference in configuration of (a), (b) and (c). While (a) and (b) have $n=4$ (number of \1 dumbbells), (c) has $n = 6$. The difference in transition energy and configuration can also be noted for size-7 (a) and (b), with $n=5$ and $n=6$, respectively. For the two defects of size 8 and size 11, the transition energy and configurations remain the same. We see that there is a direct correlation between the number of dumbbells in the \1 orientation ($n$) and transition energy for the same sized defect. For size 12, the two defects have more or less similar configurations, however the transition energy is slightly different. This might be because the defect with higher energy belongs to the M-S potential, that behaves differently from the other two potentials as discussed in \Cref{sec:potcmps}.

\begin{figure}[ht]
  \centerline{\includegraphics[width=.70\linewidth]{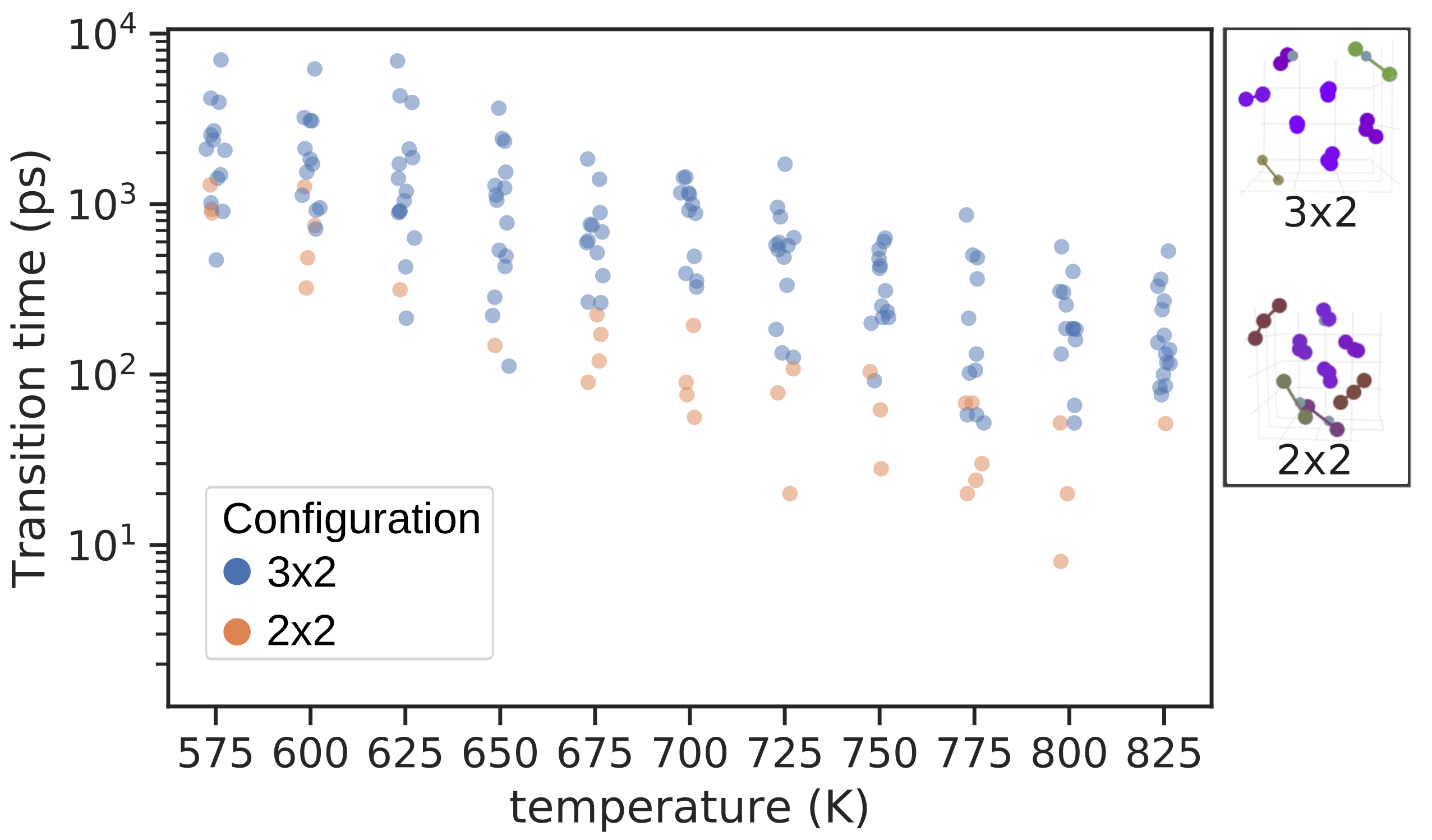}}
  \caption{\label{fig:fig8}
	Time to transition to \3 dislocation for different sample runs at various temperatures for a size 8 defect (\Cref{fig:fig4}). The transition time for the initial $2\times2$ configuration is almost always shorter. From the initial $2\times2$ configuration, the defect either changes to a more stable $3\times2$ configuration or transitions to \3 dislocation quickly.
}
\end{figure}

A defect can transition from one configuration to another as it thermally vibrates.  The two size-8 defects have initial configuration of $(2\times2+0)+4$ and in most of the sample simulations they transition to $(3\times2+2)+0$ configuration before finally transitioning to a \3 dislocation. \Cref{fig:fig8} shows the transition time as a function of temperature for all the different sample runs of the defect. We see that the transition time of the samples that remain in $2\times2$ configuration is generally lower than the samples where the configuration changes to a $3\times2$ configuration in the beginning of the simulation. Once the defect transitions to a $3\times2$ configuration it takes longer to transition while in $2\times2$ configuration it never stays as \1 loop for long. In the figure, We see that the transition time of the samples that remain $2\times2$ configuration is generally lower than the samples where the configuration changes to $3\times2$ configuration in the beginning of the simulation. The $3\times2$ configuration is present in more samples and is also a more stable configuration.

The bigger defects have a lot more possible configurations that they can
transition between. However, there are certain configurations that are very
stable and remain same for a long duration. We observe that the configurations
with $r = 0$ and $m = 0$ are not found in our dataset of 230 cascades. The
defect with size 6 and size 16 both have a few runs that transition from a
different initial configuration to a configuration with $r = 0$ and $m = 0$,
$3\times2$ and $4\times4$, respectively. These configurations are observed to
be more stable but transition to these occurs rarely.

\subsubsection{Similar transition energy of different sizes}

All of the defects in the size range of 8 to 11 have $l \times b = 3\times2$ (while $r$ varies from 0 to 1 and $m$ ranges from 2 to 4). The transition energy of all these defects are similar. For size-7, the transition energy of the defect with $(2\times2+1)$ configuration of \1 dumbbells is lower, while the one with $(3\times2+0)$ has slightly higher transition energy compared to other $3\times2$ configurations. A jump in the value of transition energy is accompanied with configurational change at size 12. The stable configuration of \1 dumbbells for size 12 defects is $l \times b = 3\times3$. For size 16 defect the transition energy as well as value of $n$ increases with possible configurations of $3\times3+2$, $4\times3+0$ or $4\times4+0$. The transition energy of sizes 19, 21 and 24 is higher than others but remains similar to each other. This agrees with the similarity in their configuration having $l \times b = 4\times3$. 

The similarity of the configuration of \1 component for a wide range of sizes
can be seen as the preference for addition of non-\1 dumbbells on the fringes
rather than having residual \1 dumbbells or changing to a lower aspect ratio
configuration. We note that the configurations with higher aspect ratio are
preferred over the ones with lower aspect ratio, e.g. we find $4\times3$ but
not $6\times2$ for 12 \1 dumbbells. Similarly, the configurations like
$4\times2$ or $5\times2$ are never observed, rather the extra dumbbells with
increase in size get added on the fringes of $3\times2$ as non-\1 dumbbells
until the size reaches to a point where $3\times3$ is possible with a few extra
non-\1 dumbbells. A low aspect ratio arrangement or addition of residual \1
dumbbells increases the fraction of \1 dumbbells on the edges of the defect.
The residual \1 dumbbells on the edges will have a high energy density which
explains the preference of configurations towards non-\1 dumbbells over
residual \1 dumbbells. \cite{rectilinear100} shows similar result using elastic
energy comparison of rectilinear and circular \1 loops. It shows that the \1
loops grow from all the four sides keeping the shape rectilinear and square (or
low aspect ratio) even for bigger sizes as opposed to \3 loops that change to
circular arrangement as the size increases.

The tendency to reduce the fraction of dumbbells on edges explains the
similarity of configurations for very similar sized defects where bigger stable
configuration is not possible. However, it does not explain that the defect
sizes that can have bigger stable configurations still stick to the smaller
configuration with more of the extra non-\1 dumbbells on the edges. This might
be due to higher formation energy required for a bigger \1 core. For instance,
a size 8 defect can not attain $3\times3$ configuration (of 9 \1 dumbbells) and
remains in $3\times2$ configuration, but size-10 or size-11 defect can have
$3\times3$ configuration with a few non-\1 dumbbells still left. However, we
only observe $3\times2$ configurations for these sizes too.  Similarly, bigger
sized defects such as size 19, 21, 24 do not have configurations that would
maximize the number of \1 dumbbells.

\subsection{Interatomic Potential and Stability\label{sec:potcmps}}

The transition energies of \1 dislocations of different sizes show similar 
values and trends for all three potentials. However, there are a few 
differences, especially in the M-S potential.

\begin{figure}[ht]
  \centerline{\includegraphics[width=.75\linewidth]{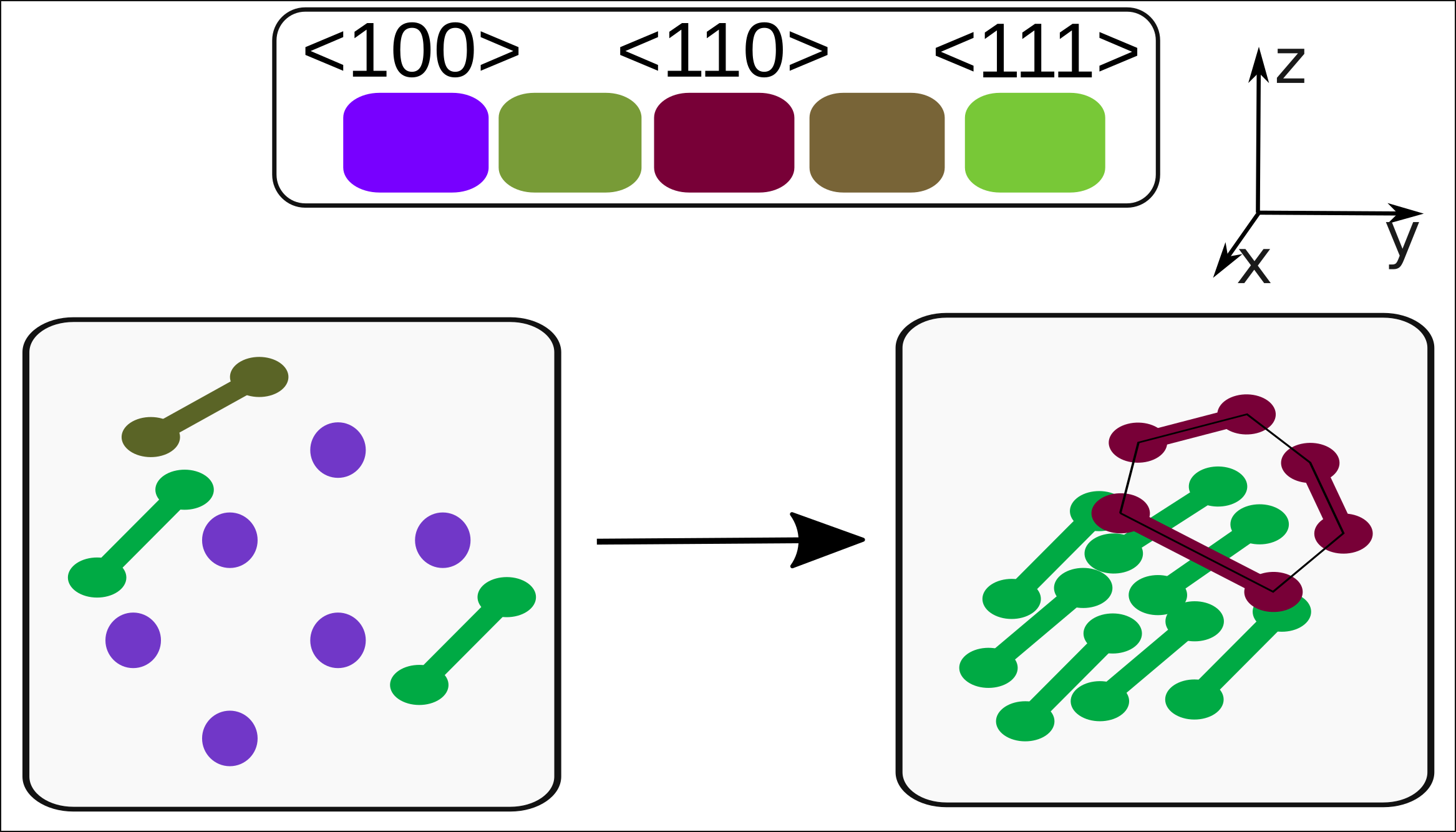}}
  \caption{\label{fig:fig9}
	The \1 dislocation with an initial configuration of 3x2 (left) and a total 
	size nine occasionally transitions to a mixed hexagonal ring (composed of 
	\2 dumbbells) + \3 dislocation morphology (right). The mixed morphology 
	defect is sessile and very stable. The dumbbells are colored according to 
	their orientation, with a color map shown in the legend. }
\end{figure}

The size-9 defect in M-S potential sometimes transitions to a mixed morphology 
defect having a component of \3 dislocation and a hexagonal ring as shown in 
\Cref{fig:fig9}. Although the transition to mixed morphology is rare, it is 
significant because once formed, this mixed configuration is very stable in M-S 
and does not transition even at very high temperatures like 2000 K. Also, 
such ring-like defects are formed relatively more often in collision cascades 
with M-S potential than with the other two potentials \cite{potcmp}. The M-S
potential has been fitted using liquid configurations in addition to perfect
crystal and point defects. For this reason, it might be more accurate about
the stability of rings in W.

Defects of size 19 onward in M-S do not transition within the limits of our
simulations, while for DND-BN, the defects up to size 24 transition. The two
defects with size 12 have same configurations but different interatomic
potentials. The one that belongs to M-S potential has higher transition energy
than the other one that belongs to the DND-BN potential (\Cref{fig:fig4}).
In \cite{BONNY2020109727}, it has been shown that the big multi-component
dislocations are highly stable in M-S, while it is not the case with other
potentials that they use. It has also been shown that the M-S potential
stabilizes \1 dislocation loops over 1/2 \3 \cite{CHEN2018141}. Our observation
also indicate that the transition energy for \1 dislocations is higher in the
M-S potential compared to the other two potentials especially for bigger sizes
in our dataset.

\section{Conclusion}\label{sec:conclude}

We show that the activation energy of a \1 dislocation to transition to \3
dislocation depends on the size and internal configuration. The bigger \1
dislocations are stable. Several defects formed in collision cascades are of
small size and might transition to \3 dislocations due to their low transition
energy. This is in agreement with the results from experiments that
show that at elevated temperatures the number of \1 loops reduce. We explored
the internal configurations of different \1 dislocations to understand the
relationship between defect size, internal configuration and factors that
affect the transition energy that are not clear when considering only the size.
We observe that many similar sized defects have similar configuration of \1
component which results in comparable transition energy for a wide band of
sizes. We show that various comparable sizes that have same \1 configurations
have same transition energy while two same sized defects with different
configurations have different transition energy. 

The similarity of \1 configurations for similar sized dislocations is due to
the preference of \1 dislocations for having non-\1 dumbbells on the fringes
rather than having residual \1 dumbbells on the corners or having a low aspect
ratio configuration with elongated spread of \1 dumbbells. A high aspect ratio
puts less dumbbells on the interface. The dumbbells in the center are known to
have lower energy density than the ones on the edges. Our results also agree
with analytical results that show that the \1 loops prefer square rectilinear shapes
\cite{rectilinear100}. No noticeable affect on the transition energy is
observed due to small changes in the fraction of non-\1 dumbbells. The
configurational dynamics show that the non-\1 dumbbells on the fringes move
around the defect. The transition occurs by rotation of dumbbells from \1 to \3
starting from the edges of the \1 component of the dislocation where energy
density is high. If a configuration has residual \1 dumbbells on the edges or
corner then the dumbbells on the edges with high energy density keep rotating
back and forth from \1 to non-\1 very often. If there are no residual \1
dumbbells and the \1 dumbbells form complete parallelogram/rhombus, the \1
dumbbell arrangement remains more stable. The movement of non-\1 dumbbells on
the fringes can also induce instability in a defect. The internal
configurations found and the transition pathways observed have good agreement
across the potentials. However, the transition energy seems to be slightly
higher for the M-S potential especially for bigger size defects.  This is in
agreement with other ab-initio and simulation results that show that the M-S
potential stabilizes \1 loops more than the \3 loops

The understanding of the correspondence of defect sizes with configurations and
trends of transition energy with configuration can be used to approximate the
transition energy values for input to higher scale models like KMC. It is
difficult to say how well the trends of transition energy with size will
generalize to very big \1 dislocations. A bigger data-set might enable
quantification of various relationships between different factors such as
transition energy, size and configurations. This study can also guide a similar
stability study for mixed dislocations morphology (having both \1 and \3
dislocations) which constitute a bigger proportion of defects formed in high
energy W collision cascades. Other crucial properties for higher scale models
include diffusion of \1 dislocations and their interactions with other defects.
The systematic studies of the defect properties for each prominent morphology
can help in higher scale modeling of the evolution of microstructure after
irradiation and for designing materials with desired properties. Other methods
such as NEB and ab-initio studies can further supplement the properties found
using dynamic MD simulations. A KMC study that accounts for transition energy,
migration energy and interactions of different morphologies and sizes accurately
can be used to match experimental results quantitatively. The presented transition
energy results and insights into stability of \1 loops form part of the input
to such a higher scale study.

\section*{Acknowledgments}
We would like to thank the supercomputing department of the Indian Plasma Research Institute for providing the high performance computing facility for carrying out the simulations.

%\section*{Data and code availability}
%The raw data required to reproduce these findings and code required to reproduce these findings are available to download from \\ https://github.com/haptork/csaransh/releases/tag/0.4 as a release for parent open source CSaransh software \cite{bhardwajcsaransh}.

\bibliographystyle{elsarticle-num} 
\bibliography{content}

\end{document}